\documentclass[twocolumn,secnumarabic,amssymb, nobibnotes, aps, prd]{revtex4-1}
\usepackage{graphicx}
\usepackage{amsmath}
\usepackage{amssymb}

\newcommand{\Otwo}{{O$_{2}$ }}
\newcommand{\Xstate}{{$X^3\Sigma_{g}^{-}$ }}
\newcommand{\Cstate}{{$C^3\Pi_{g}$ }}

\begin{document}

\title{Rotational spectroscopy with an optical centrifuge}
\author{Aleksey Korobenko}
\author{Alexander A. Milner}
\author{John W. Hepburn}
\author{Valery Milner}
\affiliation{Department of Physics \& Astronomy, University of British Columbia, 2036 Main Mall, Vancouver, BC, Canada V6T 1Z1}
\date{\today}%

\begin{abstract}
We demonstrate a new spectroscopic method for studying electronic transitions in molecules with extremely broad range of angular momentum. We employ an optical centrifuge to create narrow rotational wave packets in the ground electronic state of $^{16}$O$_2$. Using the technique of resonance-enhanced multi-photon ionization, we record the spectrum of multiple ro-vibrational transitions between \Xstate and \Cstate electronic manifolds of oxygen. Direct control of rotational excitation, extending to rotational quantum numbers as high as $N\gtrsim 120$, enables us to interpret the complex structure of rotational spectra of \Cstate beyond thermally accessible levels.
\end{abstract}

\maketitle

Ro-vibrational spectroscopy of molecules often employs the technique of resonance-enhanced multi-photon ionization (REMPI) due to the high sensitivity, spectral resolution and versatility of the latter \cite{Letokhov1987}. In a typical setup, molecules are ionized through a resonant multi-photon electronic transition with a tunable laser, and the induced ion current is measured as a function of the laser wavelength. The frequency of the observed REMPI resonances reveals the ro-vibrational spectrum of the molecule, while their strength is proportional to the population of a given initial level and the transition probability to the excited state.

One traditional approach to rotational spectroscopy involves interpreting REMPI spectra by fitting them to the theoretical model of a molecular potential, while using appropriate spectroscopic constants (e.g. rotational constant $B_{v}$) as fitting parameters. This method is hardly applicable to the study of high-energy rotational states because of the two main reasons. Firstly, populating high rotational states by means of ro-vibrational energy transfer \cite{Mullin1995} or with laser light \cite{Karczmarek1999, Li2000, Cryan2011} is rather challenging, whereas thermal excitation may require unreasonably high temperatures, e.g. $\approx$50,000 K for reaching $N>100$ in oxygen.

The second problem stems from the high number of populated rotational levels leading to high complexity of the observed spectra. For example, spin-rotation and spin-orbit level splitting both in the ground and excited states of \Otwo, together with the two-photon selection rules $\Delta J=0,\pm 1,\pm 2$ in a typically used two-photon transition $C^3\Pi_g(v'=2)\leftarrow\leftarrow X^3\Sigma_g^-(v''=0)$, would result in as many as 21 peaks for each $N$\cite{Sur1986}. For a broadly populated thermal ensemble this would lead to 21 largely overlapping branches. Finally, theoretical treatment of extremely broad range of angular momenta may require using intermediate Hund's cases or considering perturbations by other states\cite{Morrill1999}, which often result in a high level of uncertainty.

Optical centrifuge is an alternative tool for exciting molecules to extremely high rotational states by means of non-resonant laser fields \cite{Karczmarek1999, Villeneuve2000, Yuan2011, Yuan2011a}. In a recent study, we have shown that the centrifuge can be used to produce and control the so-called ``super rotor'' states - coherent rotational wave packets with ultra-high angular momentum $N$ and narrow distribution width $\Delta N \ll N$\cite{korobenko2013}. Here we utilize this unique capability of the centrifuge for the purpose of obtaining and interpreting complex REMPI spectra of oxygen super rotors ($0 < N \lesssim 120$). We excite oxygen to a narrow rotational wave packet whose center is accurately tuned across the broad range of well defined $N$ values. The centrifuge excitation is then followed by a REMPI measurement. Owing to the narrow $N$ distribution, the detected spectrum becomes significantly less congested, and identifying rotational resonances is greatly simplified.

\begin{figure*}[!htbp]
\includegraphics[width=1.95\columnwidth]{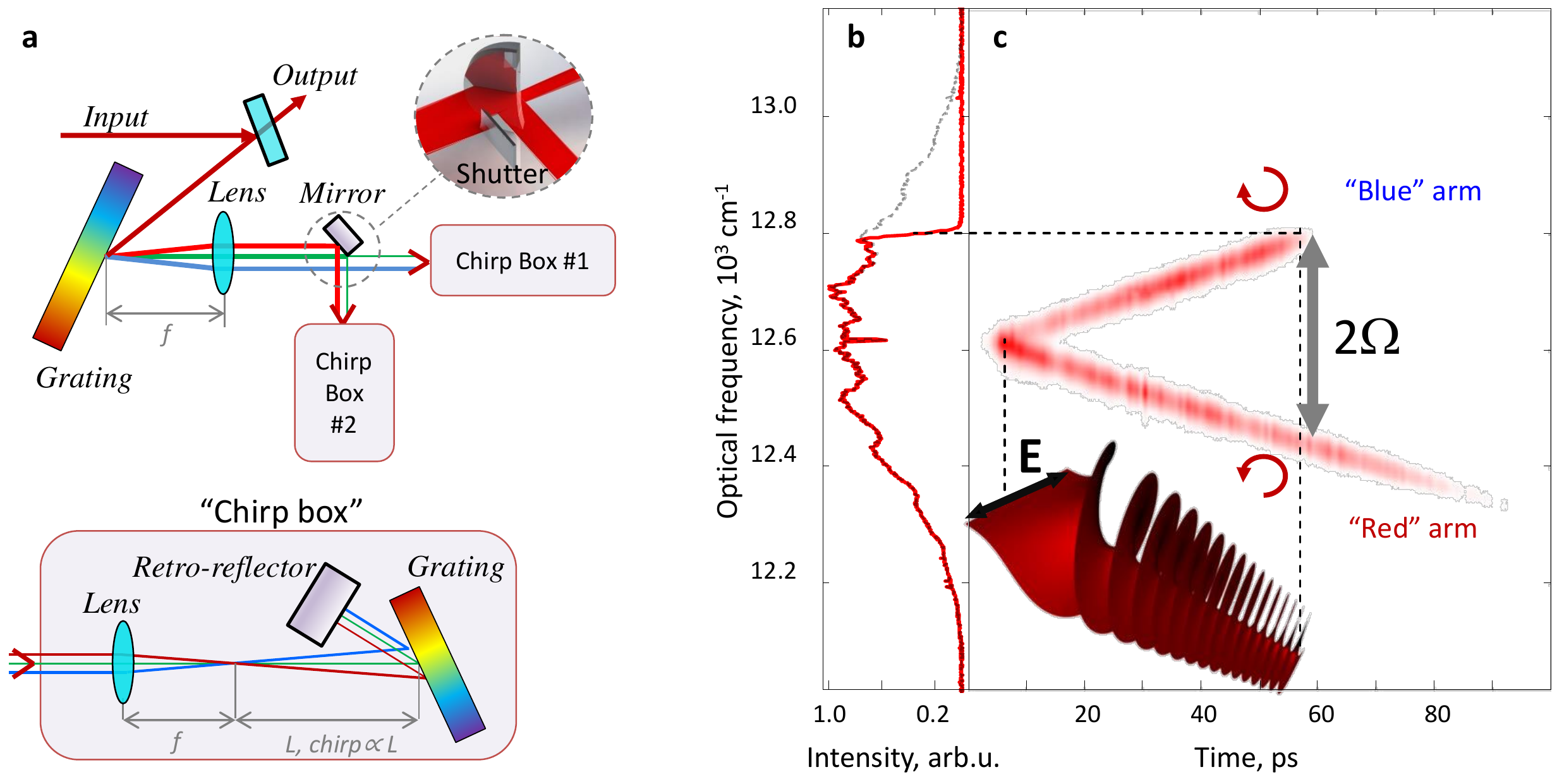}
\caption{Optical centrifuge. {\bf a.} Broadband laser pulses from a Ti:S chirped pulse amplifier (10 mJ, 35 fs, 1 KHz repetition rate) are dispersed with a diffraction grating and split in the center of the spectrum in a Fourier plane of the focusing lens into ``red'' and ``blue'' arms, whose chirps are individually controlled by two separate ``chirp boxes''. The chirp of the ``red'' arm is reversed, while that of the ``blue'' arm is left unchanged. Movable shutter on a motorized linear stage (inset) allows precise truncation of the ``blue'' arm bandwidth. {\bf b.} Spectrum of the centrifuge pulse after shaping. Solid red (dashed black) line corresponds to the truncated (full) centrifuge. {\bf c.} Cross-correlation frequency-resolved optical gating (XFROG) spectrogram of the truncated centrifuge field, schematically shown below the XFROG plot. $\Omega $ is the angular frequency at which molecules are released from the centrifuge.}
\label{fig_centrifuge}
\end{figure*}

Following the original recipe by Karczmarek et al. \cite{Karczmarek1999}, we utilize the output of an 800 nm, 35 fs (full width at half-maximum, FWHM), Ti:Sapphire regenerative amplifier. We split its spectrum in half at around the central wavelength (Fig. \ref{fig_centrifuge}{\bf a}), in a Fourier plane of a pulse shaper. Frequency chirps of equal magnitude (0.26 ps$^{-2}$) and opposite signs are applied to the ``red'' and ``blue'' arms of the centrifuge, as demonstrated by the cross-correlation frequency-resolved optical gating (XFROG) measurement (Fig. \ref{fig_centrifuge}{\bf c}). The two arms are combined with a polarizing beam splitter cube, and polarized with an opposite sense of circular polarization. Optical interference of the two circularly polarized frequency-chirped laser fields results in a pulse with rotating linear polarization (inset to Fig. \ref{fig_centrifuge}\textbf{c}). Because of the anisotropic polarizability, molecular axes line up along the axis of laser polarization, and then follow it adiabatically as the plane of polarization rotates with increasing angular frequency. Given the available spectral bandwidth, the accelerating centrifuge can reach angular frequencies on the order of 10 THz, which in the case of oxygen corresponds to the rotational quantum number $N \approx 119$.

As we have demonstrated in \cite{korobenko2013}, truncating the spectrum of the centrifuge in a Fourier plane of the pulse shaper by a movable shutter (inset to Fig. \ref{fig_centrifuge}{\bf a}) enables accurate control of the rotational state of the centrifuged molecules. Characterizing the centrifuge field with XFROG allowed us to calibrate the final rotation speed of the centrifuge, and hence the corresponding molecular angular momentum, as a function of the shutter position.

\begin{figure}[!htbp]
\includegraphics[width=1\columnwidth]{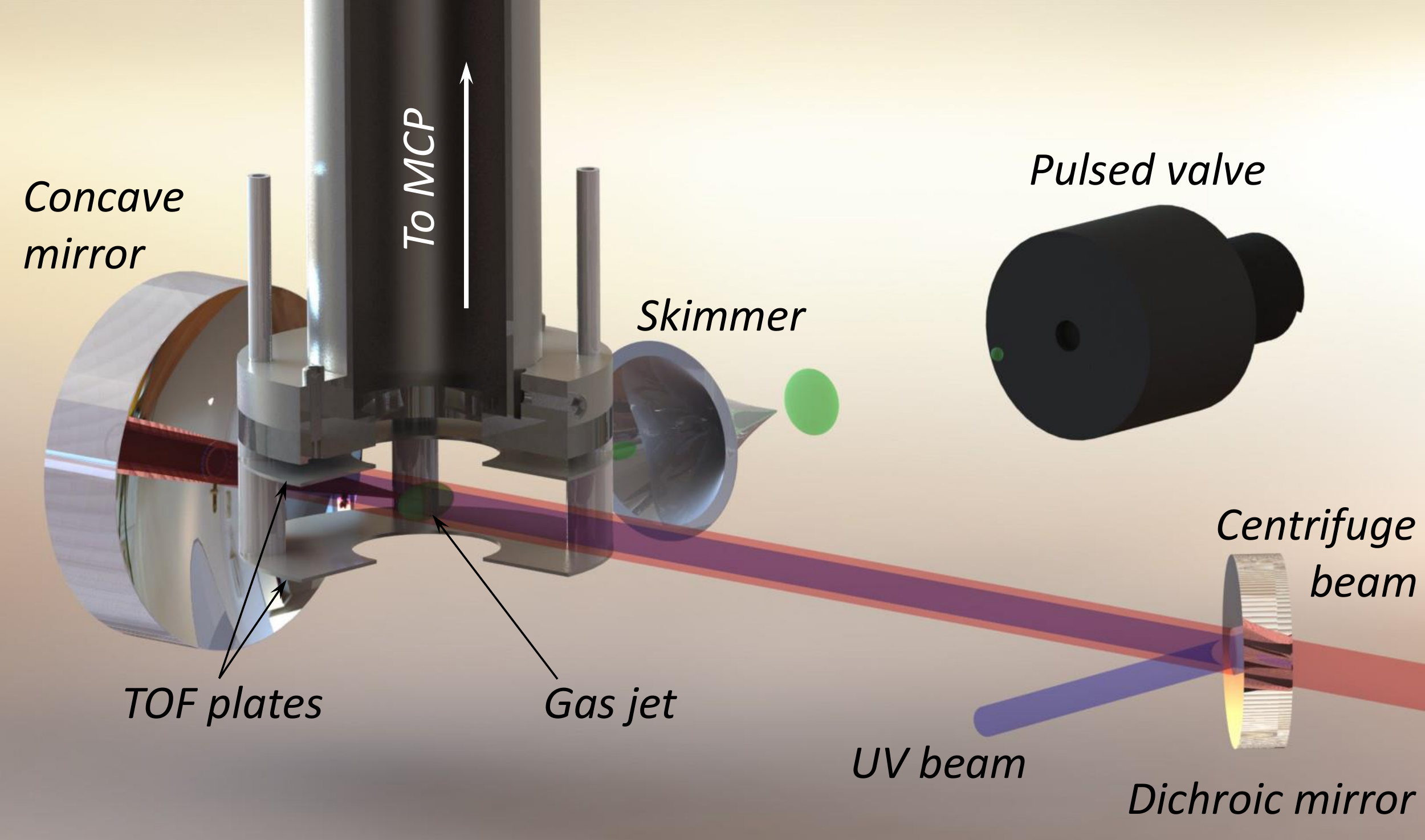}
\caption{ Detection setup. Centrifuge beam is combined with a tunable UV laser pulse and focused inside a vacuum chamber on a supersonically expanded oxygen jet between the charged plates of a time-of-flight (TOF) mass spectrometer. The ionization rate is measured with a multi-channel plate (MCP).}
\label{fig_setup}
\end{figure}

\begin{figure*}[!htbp]
\includegraphics[width=1.95\columnwidth]{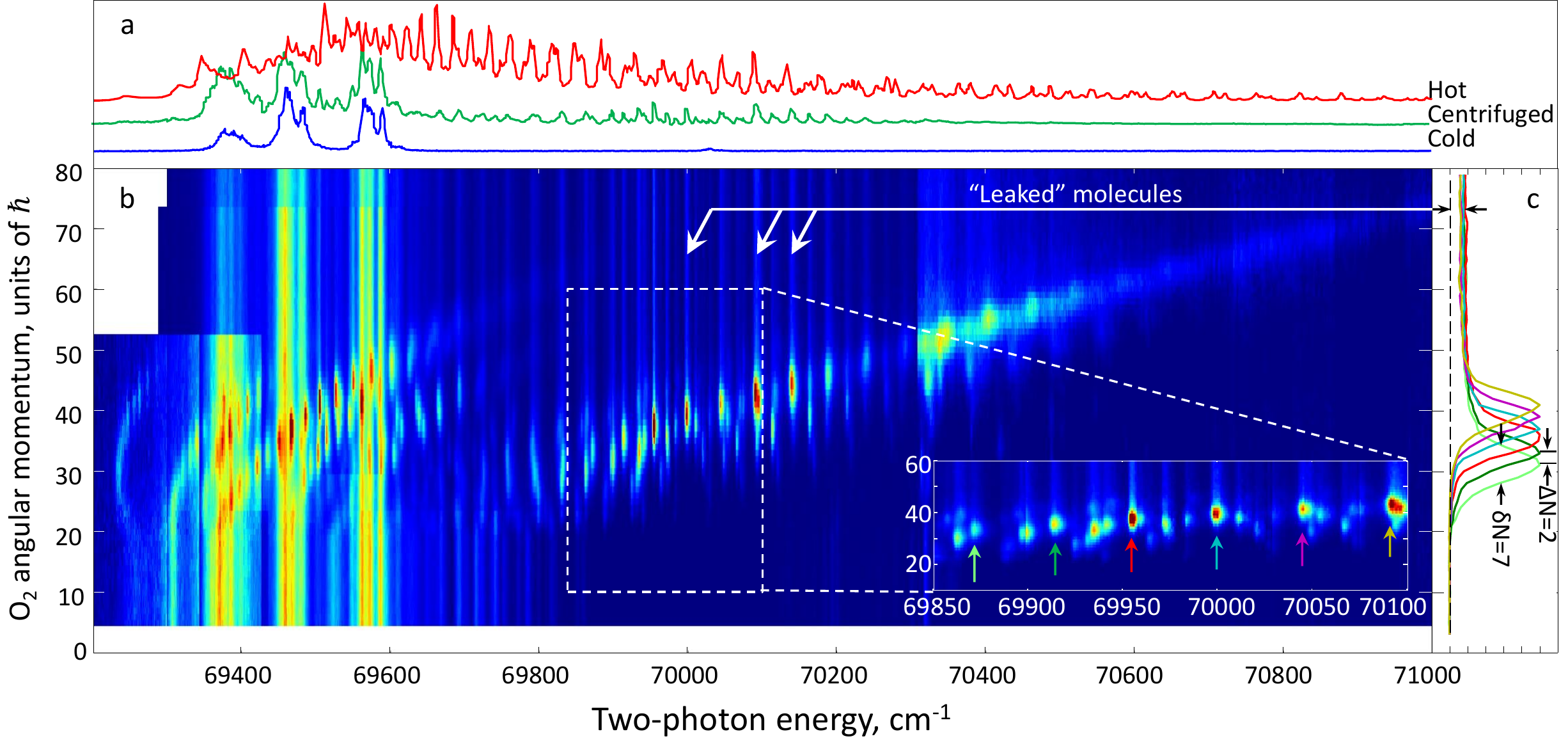}
\caption{2D REMPI spectrogram for a linearly polarized probe. {\bf a.} Experimental spectra of cold (10~K, blue) and centrifuged molecules (green), along with a simulated spectrum of a ``hot'' thermal ensemble (3000~K, red) calculated with \texttt{PGOPHER} software\cite{Pgopher}. {\bf b.} Ion signal as a function of the probe laser wavelength and molecular angular momentum defined by the centrifuge final rotation speed. Different areas of the 2D plot were measured with different sensitivities and probe intensities and are displayed with different color scales to compensate for the broad dynamic range of the data. {\bf c.} Vertical cross-sections of several consequent peaks from one particular branch, shown in the inset to {\bf b}. The peaks are regularly separated with a distance of $\Delta N=2$ reflecting $^{16}O_2$ nuclear spin statistics.}
\label{fig_overall}
\end{figure*}

REMPI detection was carried out using narrowband nanosecond probe pulses tunable from 279 nm to 288 nm (0.1 cm$^{-1}$ linewidth, 500 $\mu $J, 50 Hz repetition rate). Probe beam was combined with a centrifuge beam on a dichroic mirror (Fig. \ref{fig_setup}), and focused with a 35 mm focal length spherical aluminium mirror on a supersonically expanded molecular jet passing between the charged plates of the time-of-flight (TOF) spectrometer. The jet was generated by an Even-Lavie pulsed valve ($25~\mu\mathrm{s}$ opening time, $150~\mu\mathrm{m}$ nozzle diameter) located $20~\mathrm{cm}$ away from the detection region. Ion current was detected with a microchannel plate (MCP) detector. Initial rotational temperature of the sample, extracted from the REMPI spectrum taken without the centrifuge field (Fig. \ref{fig_overall}{\bf a}), was about 10 K.

The main result of this work is shown in Figure \ref{fig_overall}, where the detected ion count is plotted against the probe wavelength (horizontal axis) and the final rotation speed $\Omega $ of the truncated centrifuge (vertical axis). The latter is expressed in terms of the angular momentum $N$ of an oxygen molecule rotating with the angular frequency $\Omega $, according to:
$$\Omega=2\pi c \left[ E(N)-E(N-1) \right],$$
$$E(N)=BN(N+1)-DN^2(N+1)^2,$$
where $E$ is the energy of state $|N\rangle$, $c$ is the speed of light in vacuum, $B=1.438$ cm$^{-1}$ and $D=4.839\times 10^{-6}$ cm$^{-1}$ \cite{NIST}. The validity of Dunham expansion of rotational energy to second order in $N(N+1)$ at extremely high values of $N$ has been demonstrated in our previous study\cite{korobenko2013}.

Each peak in the two-dimensional REMPI spectrogram of Fig. \ref{fig_overall}{\bf b} corresponds to a two-photon transition between a rotational level in the electronic ground state, \Xstate, and a rotational level of \Cstate. Finite horizontal width of the observed peaks stems from the finite transition linewidth (as in conventional ``1D REMPI'' detection), whereas finite vertical spread reflects the narrow width of the excited rotational wave packet created by the centrifuge.

The complexity of the two-photon absorption line structure in rotationally hot oxygen gas is illustrated by red and green lines in Fig.\ref{fig_overall}{\bf a} which correspond to the hot thermal ensemble (simulated numerically) and the ensemble of centrifuged molecules (experimentally observed 2D spectrogram integrated along its vertical dimension), respectively. In sharp contrast to conventional 1D REMPI spectroscopy, controlled centrifuge spinning offers direct assignment of rotational quantum numbers to the observed REMPI peaks, as well as significantly better peak separation due to their distribution along the added second dimension.

Vertical traces originating from bright resonance peaks in Figure \ref{fig_overall}{\bf b} (examples are marked with white arrows) correspond to molecules which ``leaked out'' of the weakened centrifuge potential before reaching the terminal angular frequency of the centrifuge. After escaping the centrifuge, these molecules continue their free rotation while the trap is accelerating further. The three bright vertical stripes reproduce the initial cold beam spectrum (blue line in Fig. \ref{fig_overall}{\bf a}) and correspond to the molecules which were not trapped by the centrifuge. The width of the final rotational wave packets can be readily extracted as $\delta N\approx 7$ (FWHM), from the vertical cross sections, shown in Fig. \ref{fig_overall}{\bf c}. Here, we detected rotational states with $N$ as high as $\sim 80$. Rotational line broadening above $N\approx 60$ can be attributed to the increasing Rydberg-valence interaction (governed by the Frank-Condon overlap with the continuum wavefunctions) similarly to the previously observed rotational broadening in the lower vibrational states ($v'=0,1$) of the excited potential\cite{Sur1986}.

\begin{figure*}[!htbp]
\includegraphics[width=1.95\columnwidth]{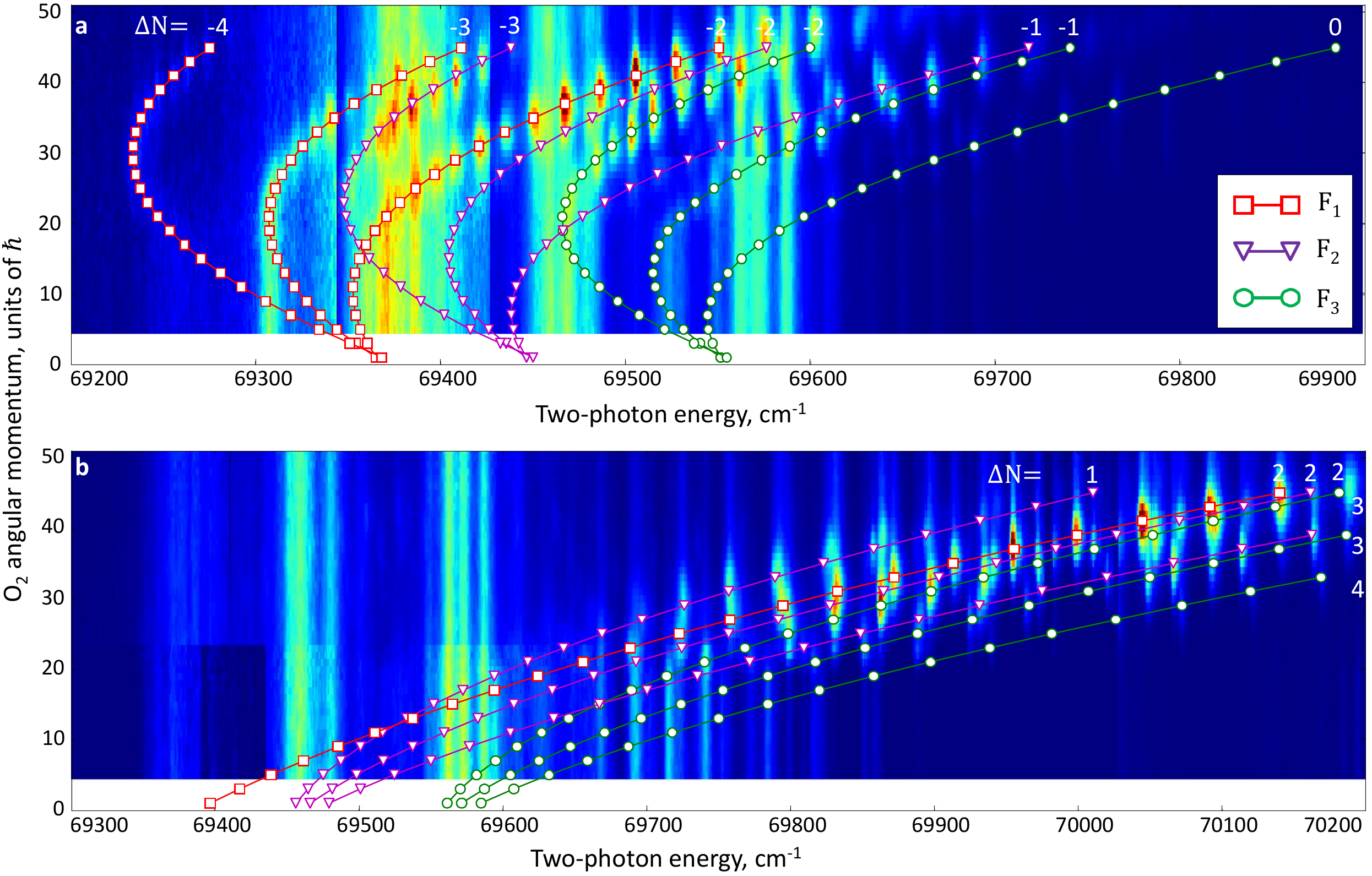}
\caption{2D REMPI spectrogram for a circularly polarized probe. Electric field vector is counter-rotating ({\bf a}) and co-rotating ({\bf b}) with the centrifuged molecules. The directionality of laser-induced rotation results in the sensitivity of the measured signal to the handedness of probe polarization. The results of fitting the data to the theoretical model are shown with colored lines and markers for different branches and resonances, respectively (see text for details). Branch nomenclature is the same as in \citep{Sur1986}.}
\label{fig_fits}
\end{figure*}
One can see that the peaks in Fig.\ref{fig_overall} are grouped in regular patterns, resembling Fortrat parabolas corresponding to different rotational branches. Within a single branch, the center of each consecutive resonant peak is shifted by $\Delta N =2$ (Fig. \ref{fig_overall}{\bf c}), reflecting the smallest step in the rotational ladder climbing executed by the centrifuge. Circularly polarized light can be used to further simplify the spectrum. As shown in Fig. \ref{fig_fits}, the signal strength of different rotational branches depends on the handedness of probe polarization. This is due to the highly non-uniform population distribution among the magnetic sub-levels in the centrifuged wave packet, with most of the population concentrated at $M_J\simeq J$ (or $M_J\simeq-J$) \cite{Karczmarek1999}.

\begin{figure}[!htbp]
\includegraphics[width=0.95\columnwidth]{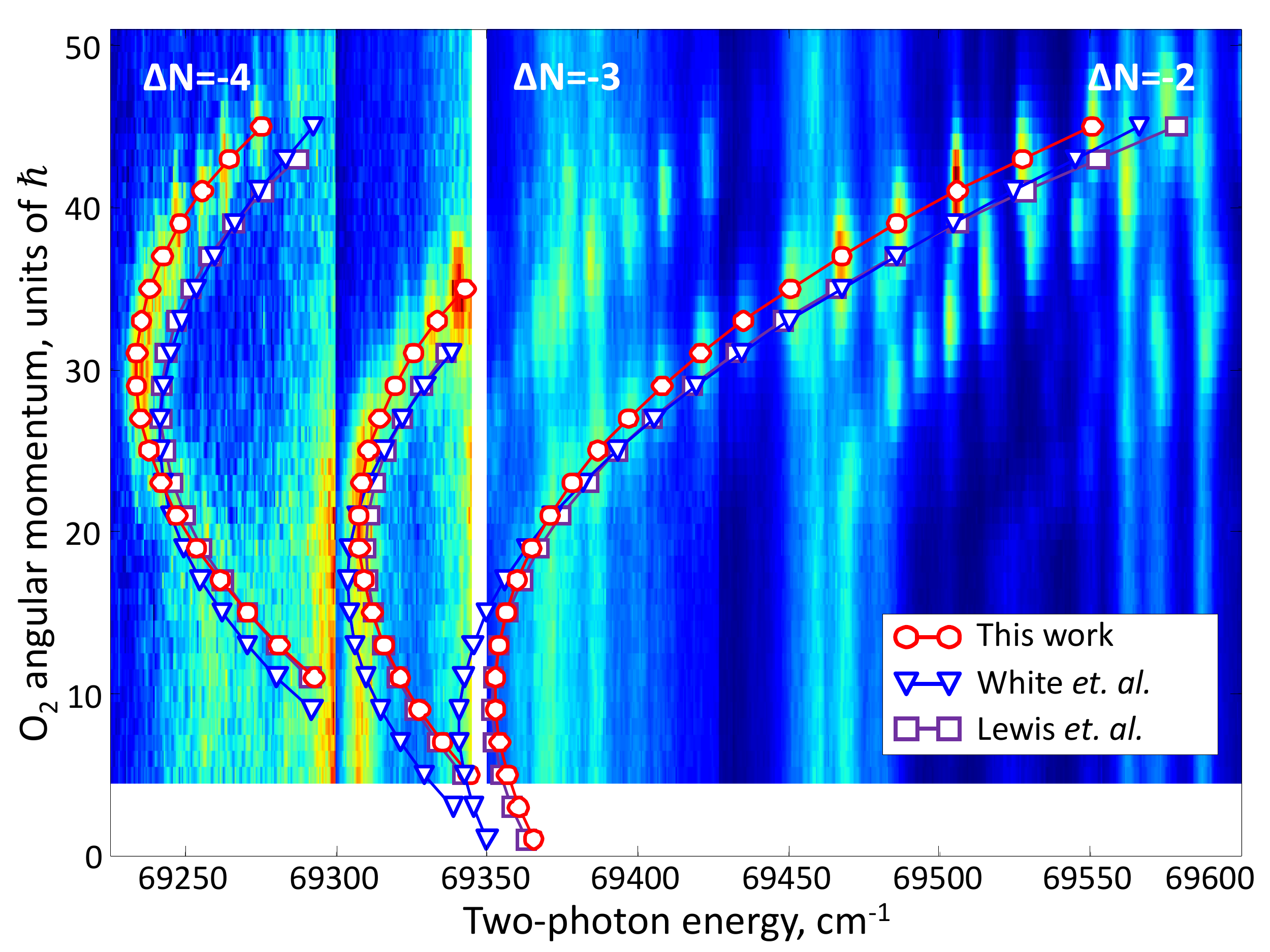}
\caption{Comparison of the observed REMPI data for the perturbed $F_1$ spin-orbit component with the calculations based on spectroscopic constants from this work (red circles), White \textit{et al.}\cite{white2004} (blue triangles) and Lewis \textit{et al.}\cite{Lewis1999} (purple squares).}
\label{fig_comparison}
\end{figure}

To identify different rotational branches, we use three sets of spectroscopic constants (for $F_1,F_2$ and $F_3$ spin-orbit components of the excited state, respectively) from the previous studies on thermally excited ensembles\cite{Lewis1999, white2004}. These constants are listed in Table \ref{tab_constants}. For $F_2$ and $F_3$ components, our results are well described by the constants provided by Lewis \textit{et al.}\cite{Lewis1999}. On the other hand, the observed $F_1$ peaks do not agree well with the suggested numerical values ($\nu_0=69366$ cm$^{-1}$ and $B_0=1.6$ cm$^{-1}$), as shown in Figure \ref{fig_comparison}. This can be attributed to the complexity of the broadened and highly overlapping structure of $F_1$ lines, which makes it hard to interpret and fit the data from a thermally populated ensemble. Centrifuge spectroscopy enabled us to correct the values of $F_1$ spectroscopic parameters (Table \ref{tab_constants}) by performing the fit of the most pronounced $\Delta N=-2$ branch (Fig. \ref{fig_comparison}).

\begin{table}
\begin{tabular}{|c|c|c|c|}
\hline
{\bf Spin-orbit branch} & $\mathbf{\nu_0}${\bf, cm}$^\mathbf{-1}$ & $\mathbf{B_0}${\bf , cm}$^\mathbf{-1}$ & {\bf D, cm}$^\mathbf{-1}$\\
\hline
$^3\Pi_0(F_1)$ & 69375 & 1.585 & $2.5\times 10^{-7}$\\
\hline
$^3\Pi_1(F_2)$ & 69445 & 1.648 & $1.0\times 10^{-5}$\\
\hline
$^3\Pi_2(F_3)$ & 69550 & 1.685 & $1.3\times 10^{-5}$\\
\hline
\end{tabular}
\caption{Spectroscopic constants used to fit the data in Fig.\ref{fig_fits}}
\label{tab_constants}
\end{table}

\begin{figure}[!htbp]
\includegraphics[width=0.95\columnwidth]{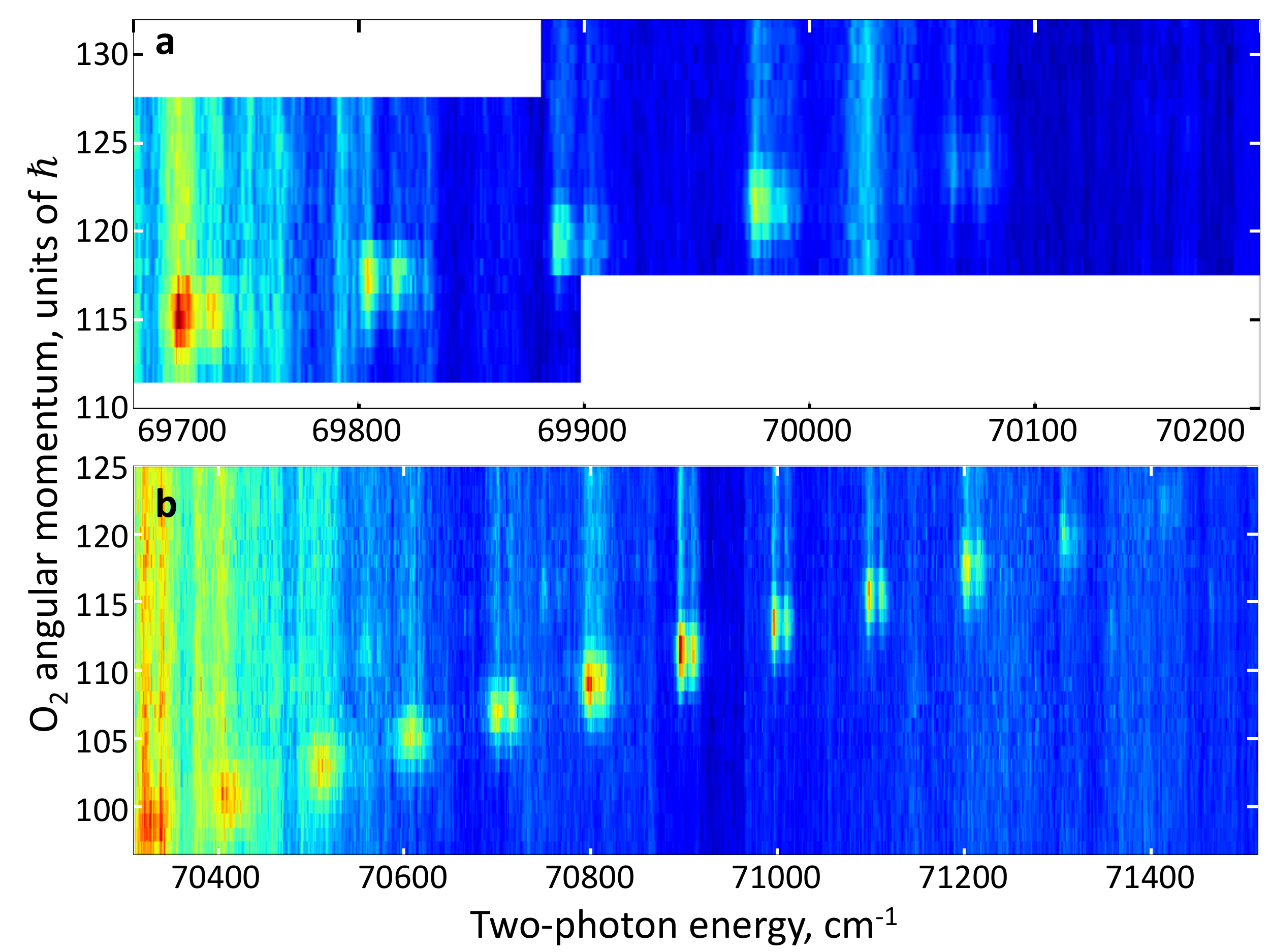}
\caption{Ultra high-$N$ resonances. Probe polarization in {\bf a} and {\bf b} is the same as in Fig. \ref{fig_fits}}.
\label{fig_high}
\end{figure}

In addition to the identified branches of $C^3\Pi_g(v'=2)\leftarrow\leftarrow X^3\Sigma_g^-(v''=0)$, washing out at $N\gtrsim 80$, we observed well resolved ultra-high rotational lines with probe polarizations both counter-rotating (Fig. \ref{fig_high}{\bf a}, branch ``a'') and co-rotating (Fig. \ref{fig_high}{\bf b}, branch ``b'') with respect to molecular motion.
The corresponding rotational numbers exceeded 100 and reached 123 for the highest detected transition. Line separation in branch ``a'' was $84~\mathrm{cm}^{-1}$ at $N=117$, and $102~\mathrm{cm}^{-1}$ in branch ``b'' at $N=113$. The lines appear in pairs, with a splitting of about $12~\mathrm{cm}^{-1}$ for both branches, which can be attributed to $\Lambda$-doubling. Strong linewidth dependence on $N$ was detected in branch ``b'': the doublet structure could not be resolved at $N=103$, whereas it was easily identified at $N=115$, where the linewidth of each individual resonance was found to be equal $7~\mathrm{cm}^{-1}$. While ``b'' branch resonances can still be attributed to $v'=2\leftarrow\leftarrow v''=0$ transition, branch ``a'' lies $\sim 1000~\mathrm{cm}^{-1}$ lower, pointing at its possible origin from a different vibrational state.

In conclusion, we have demonstrated a new spectroscopic method for studying the rotational structure of electronic transitions in molecules. The method is based on controlled molecular spinning with an optical centrifuge. We have applied this technique to $C^3\Pi_g(v'=2)\leftarrow\leftarrow X^3\Sigma_g^-(v''=0)$ in \Otwo, and showed an agreement with previously known spectroscopic constants for $F_2$ and $F_3$ spin-orbit components. In the case of strongly perturbed $F_1$ components, higher resolution of the implemented method enabled us to improve the previously known spectroscopic constants. We also observed narrow ($\sim 7~\mathrm{cm}^{-1}$) resonances from ultra-high rotational states with $N>100$ with highly non-monotonic linewidth dependence. Some of these resonances were attributed to a different excited vibrational state, which could not be resolved in previous studies at low $N$.

This work has been supported by the CFI, BCKDF and NSERC, and carried out under the auspices of the Center for Research on Ultra-Cold Systems (CRUCS). We gratefully acknowledge stimulating discussions with R. Krems, V. Petrovic,  M. Shapiro and E. Grant.


%

\end{document}